\documentclass[multphys,vecphys]{svmult}

\usepackage{makeidx}         
\usepackage{graphicx}        
\usepackage{multicol}        
\usepackage[bottom]{footmisc}

\makeindex             

\begin{document}

\title*{Homogeneous comparison of directly detected planet 
candidates: GQ Lup, 2M1207, AB Pic}
\author{Ralph Neuh\"auser\inst{1} }
\institute{Astrophysikalisches Institut, Schillerg\"asschen 2-3,
D-07745 Jena, Germany, rne@astro.uni-jena.de}
%
%
\maketitle

{\bf Abstract.} We compile the observational evidence for the three recently
presented planet candidates imaged directly and derive in a 
homogeneous way their temperatures and masses.
For both AB Pic b and 2M1207 b, we derive a larger 
temperature range than in Chauvin et al. (2004, 2005b).
AB Pic b appears to be quite similar as GQ Lup b, but older.
According to the Tucson and Lyon models, all three companions could either
be planets or brown dwarfs. According to the Wuchterl formation model,
the masses seem to be below the D burning limit. We discuss whether
the three companions can be classified as planets, and whether 
the three systems are gravitationally bound and long-term stable.

\section{Introduction: Direct imaging of exo-planets}

Direct imaging of planets around other stars is difficult because
of the large dynamic range between faint planets very close
to much brighter primary stars.
Few Myr young planets and young sub-stellar companions in general
including both brown dwarfs are much brighter
than Gyr old sub-stellar objects because of on-going contraction
and possibly accretion (e.g. Burrows et al. 1997;
Wuchterl \& Tscharnuter 2003).

Below, we will compile the observational evidence
published for the three currently discussed exo-planets detected
directly, namely around GQ Lup (Neuh\"auser et al. 2005a; henceforth N05a),
2M1207 (Chauvin et al. 2005a; henceforth Ch05a),
and AB Pic (Chauvin et al. 2005b; henceforth Ch05b).
From the published observables, we derive in a homogeneous way
the parameters needed for placement in the H-R diagram,
i.e. luminosity and temperature.
Then, we compare the loci of these three planet candidates with 
different model tracks to determine the masses.

\section{Observational evidence: Three candidates}

N05a presented astrometric and spectroscopic evidence for a sub-stellar
companion around the well-known classical T~Tauri star GQ Lup, 
for which also radius and gravity could be determined.
Chauvin et al. (2004) presented a companion candidate near
2MASSWJ 1207334-393254 (or 2M1207 for short), JHK imaging and
a low-resolution, low S/N spectrum (both with AO at the VLT),
which still needed astrometric confirmation.
Schneider et al. (2004) also detected the companion candidate
using the HST/Nicmos a few weeks later, too early for
astrometric confirmation ($2~\sigma$ only).
Then, Ch05a published the astrometric confirmation for the
two objects (companion candidate and primary object) to
have the same proper motion.
Also very recently, Ch05b presented evidence for another
possibly planetary-mass companion around yet another young
nearby star, namely AB Pic.

The directly observed parameters are presented in table 1,
keeping the preliminary designations ({\em A} for the primary object,
{\em b} for the companion, always regarded as a planet candidate).
We also would like to note that both GQ Lup A and AB Pic A are normal stars,
while 2M1207 A is a brown dwarf.

\begin{table*}
\begin{tabular}{l|llllll}
\multicolumn{7}{c}{\bf Table 1. Observables published} \\ \hline
Object   & Spec  & J     & H     & K     & L or L' & distance \\
         & type  & [mag] & [mag] & [mag] & [mag]   & [pc]     \\ \hline
GQ Lup A & K7    & 8.605 (21) & 7.702 (33) & 7.096 (20) & 6.05 (13) & $140 \pm 50$ \\
GQ Lup b & M9-L4 &       &    & 13.10 (15) & 11.7  (3)  & \\ \hline
2M1207 A & M8    & 12.995 (26) &12.388 (27)& 11.945 (26)& 11.38 (10) & $70 \pm 20$ (*) \\
2M1207 b & L5-9  & $\ge 18.5$ & 18.09 (21) & 16.93 (11) & 15.28 (14) & \\ \hline
AB Pic A & K2    & 7.576 (24) & 7.088 (21) & 6.981 (24) & & $47.3 \pm 1.8$ \\
AB Pic b & L0-3  & 16.18 (10) & 14.69 (10) & 14.14 (8)  & & \\ \hline
\end{tabular}
\\
Note: Numbers in brackets are error margins on last digits. 
(*) Mamajek 2005 give $53 \pm 6$ pc for 2M1207 A, within the Ch05a error. \\
Ref.: N05a, Ch05a, Ch05b, 2MASS, Jayawardhana et al. 2003,
Chauvin et al. 2004.
\end{table*}

\section{Derived parameters}

Based on the directly observable parameters listed in table 1,
we can now homogeneously derive some other parameters,
which are not observable directly.
Those other parameters are in particular luminosity and temperature,
which are neccessary for placement into the H-R diagramm.

The derivation of temperature from the spectral type also
needs the gravity as input (see e.g. Gorlova et al. 2003).
For none of the six objects involved, the gravity is
measured directly by high-resolution spectra; only for GQ Lup b,
there is a measurement (from a low-resolution spectrum, R$\simeq 700$).
Only one of the six objects is already on the
zero-age main-sequence, namely AB Pic A, so that we can
assume dwarf gravity. GQ Lup A is a pre-MS star, 2M1207 A a brown
dwarf, and the other companions are sub-stellar and, hence,
above the main sequence, probably intermediate between dwarfs and giants.
Hence, it is best to derive the full possible range in temperature,
given several different spectral type to temperature scales.
We list the temperature ranges for all available scales in table 2 -
together with the bolometric corrections used to estimate
the luminosities, which are also given, as well as absolute
K-band magnitudes.

\begin{table*}
\begin{tabular}{l|l|ll|l}
\multicolumn{5}{c}{\bf Table 2. Derived parameters for sub-stellar objects involved} \\ \hline
Temp. scale below & GQ Lup b   & 2M1207A    & 2M1207 b  & AB Pic b  \\ \hline
~~~~~~~~~~~~~~Spectral type:   & M9-L4      & M8          & L5-9      & L0-3      \\ \hline
Luhmann 1999 (a) & $\le 2550$ & $\le 2720$  & n/a       & $< 2550$  \\
Reid et al. 1999  & 2100-1800  & $\sim 2200$ & 2000-1850 & 2000-1850 \\
Kirkpatrick et al. 2000 & 2050-1650  & n/a         & 2000-1750 & 2000-1750 \\
Basri et al. 2000     & 2500-1850  & $\sim 2500$ & 1750-1600 & 2250-1950 \\
Stephens et al. 2001  & 2320-1820  & $\sim 2400$ & 1720-1320 & 2220-1920 \\
Leggett et al. 2002 (b) & 2500-1700  & $\sim 2200$ & 1650-1150 & 2350-1650 \\
Burgasser et al. 2002  & 2300-1740  & $\sim 2400$ & 1625-1170 & 2190-1850 \\
Dahn et al. 2002       & 2500-1900  & $\sim 2550$ & 1900-1300 & 2400-1950 \\
Nakajima et al. 2004   & 2520-1830  & $\sim 2650$ & 1690-1140 & 2380-1970 \\
Golimowski et al. 2004 & 2400-1600  & $\sim 2500$ & 1950-1100 & 2400-1600 \\ \hline
mean & $2060 \pm 180$ & $2425 \pm 160$ & $1590 \pm 280$ & $2040 \pm 160$ \\ \hline
range      & 2520-1600  & 2650-2200   & 2000-1100 & 2400-1600 \\ \hline \hline
M$_{\rm K}$ [mag]  & $7.37 \pm 0.96$ & $7.72 \pm 0.66$ & $12.70 \pm 0.75$ & $10.77 \pm 0.14$ \\ \hline
B.C.$_{\rm K}$ (*) & $3.3 \pm 0.1$   & $3.1 \pm 0.1$   & $3.25 \pm 0.1$   & $3.3 \pm 0.1$ \\ \hline
$\log$ L$_{\rm bol}/$L$_{\odot}$ & $-2.37 \pm 0.41$ & $-2.43 \pm 0.20$ & $-4.49 \pm 0.34$ & $-3.730 \pm 0.039$ \\ \hline
\end{tabular}
\\
Remarks: (a) Luhman 1999 intermediate scale; (b) compilation in Leggett et al. (2002);
n/a for not applicable; all temperatures are given in [K], (*) bolometric correction 
B.C.$_{\rm K}$ in [mag] for the K-band according to Golimowski et al. (2004).
\end{table*}

The temperature mean and range for GQ Lup b is almost identical to
the one given in N05a (mean $\sim 2050$ K, range 1600 to 2500 K),
where a few scales listed in table 2 here were not included.

Chauvin et al. (2004) derive the age of 2M1207 A by assuming it to
be co-eval with the mean TWA age and then assume that 2M1207 b has
the same distance and age.
For 2M1207 b, Chauvin et al. (2004) give a temperature of
only $1250 \pm 200$ K, obtained from the absolute magnitudes in
H, K, and L' with Chabrier et al. (2000)
and Baraffe et al. (2002); apparently, this temperature range is
obtained from {\em http://perso.ens-lyon.fr/isabelle.baraffe/DUSTY00} models
for 10 Myrs, roughly the age of the TW Hya association. 
They also obtain 1000 to 1600 K from Burrows et al. (1997)
for 70 pc and 5-10 Myrs age.
Hence, they have obtained the temperature from uncertain
models tracks and an assumed distance and age, and not from 
converting the observed (distance-independant)
spectral type to a temperature.
Our temperature range is larger and its upper limit is shifted 
to higher values compared to Chauvin et al. (2004).
The situation is similar for AB Pic b, for which we obtain
a temperature of $2040 \pm 160$ K from its spectral type
and considering all scales (table 2).
Ch05b, however, only use the models by Burrows et al. (1997)
yielding 1513 to 1856 K and
Chabrier et al. (2000) and Baraffe et al. (2002) giving 1594 to 1764 K.
Hence, one could conclude that the models underestimate the temperature.

On the other hand, if one gives a correct absolute magnitude 
(or luminosity) as input (assuming a correct distance),
and also taking into account that the Lyon models used in Ch05b were
previously found to {\em under}estimate the radii (Mohanty et al. 2004),
one would expect that the resulting temperature is an {\em over}estimate.
This shows that the determination of the temperature should be
done with great care under full consideration of the young
age and, hence, low gravity of the involved objects.

\section{Mass determination by model tracks}

Once temperatures, absolute magnitudes, and luminosities are determined homogeneously,
we can derive the masses of the objects, see table 3.

\begin{table*}
\begin{tabular}{l|l|l|llll}
\multicolumn{7}{c}{{\bf Table 3. Masses of sub-stellar objects involved} (masses in [M$_{\rm jup}$])} \\ \hline
Model     & Figure   & Input           & \multicolumn{4}{c}{Object} \\
Reference & used     & parameters      & GQ Lup b    & 2M1207 A & 2M1207 b & AB Pic b \\ \hline
          &        & (age used:)     & 1-2 Myr     & 5-12 Myr & 5-12 Myr & 30-40 Myr \\ \hline
\multicolumn{7}{c}{masses derived from temperatures and ages:} \\ \hline
Burrows et al. 1997  & Fig. 9/10 & T \& age & 4-15     & 14-25    & 4-14     & 13-25 \\
Chabrier et al. 2000 & Fig. 2 & T \& age (a)& $\le 20$ & 15-25    & $\le 15$ & 15-30 \\
Baraffe et al. 2002  & Fig. 2 & T \& age (b)& 3-16     & 15-25    & 2-12     & 12-50 \\
Baraffe et al. 2002  & Fig. 3 & T \& age    & 5-30     & 20-45    & $\le 20$ & 15-30 \\
Wuchterl model       & (f)    & T \& age    & 1-3      & 1-5      & n/a (c)  & n/a (c) \\ \hline
\multicolumn{7}{c}{masses derived from luminosities and ages:} \\ \hline
Burrows et al. 1997  & Fig. 7 & L \& age    & 12-32    & 20-30    & 2-10     & 14-15 \\
Baraffe et al. 2002  & Fig. 2 & L \& age    & 12-42    & 12-30    & 2-5      & $\sim 20$ \\
Baraffe et al. 2002  & Fig. 3 & L \& age    & 10-30    & 10-50    & n/a (c)  & n/a (c)   \\
Baraffe et al. 2002  & (b)    & L \& age    & 18-50    & 25-60    & 3-6      & 11-18  \\
Wuchterl model       & (f)    & L \& age    & 1-3     & 1-5      & n/a (c)  & n/a (c) \\ \hline
\multicolumn{7}{c}{masses derived from luminosities and temperatures (H-R diagram):} \\ \hline
Burrows et al. 1997  & Fig. 11 & L \& T     & $\le 15$ & $\le 25$ & 2-70 (d) & 2-70 (d) \\
Baraffe et al. 2002  & Fig. 1  & L \& T     & $\le 20$ & $\le 20$ & n/a (d)  & n/a (d)  \\
Baraffe et al. 2002  & Fig. 6  & L \& T     & $\le 30$ & 10-35    & n/a (c)  & n/a (e)   \\
Wuchterl model       & (f)     & L \& T     & 1-3      & 1-5      & n/a (c)  & n/a (c) \\ \hline
\end{tabular}
\\
Remarks: n/a for not applicable,
(a) Similar for Dusty, Cond, and NextGen,
(b) see also http://perso.ens-lyon.fr/isabelle.baraffe/DUSTY00~models,
(c) outside of range plotted or calculated,
(d) full mass range possible; for additional contraint of assumed age, i.e. to be located
on the correct isochrone, the mass would be $\le 20$~M$_{\rm jup}$,
(e) $\le 60$~M$_{\rm jup}$ from L \& T in Fig. 8,
(f) Fig. 4 in N05a.
\end{table*}

Table 3 shows that for all three planet candidates,
there is a large mass range when employing the full possible range
of luminosities, temperatures, and age,
at least when using the Lyon or Tucson models.
For 2M1207 b, those models tend to give masses below $\sim 20$~M$_{\rm jup}$
from luminosities, temperatures, and age,
but higher masses are not excluded.

For young objects as the ones considered here, one has to take into
account the formation, i.e. initial conditions matter, so that models
starting with an assumed internal structure are highly uncertain.
Stevenson (1982) wrote about such collapse calculations:
{\em Although all these calculations may reliably represent the degenerate cooling
phase, they cannot be expected to provide accurate information on the first
$10^{5}$ to $10^{8}$ years of evolution because of the artificiality of an initially
adiabatic, homolously contracting state.} \\
Baraffe et al. (2002) also wrote that {\em assinging an age} (or mass) {\em to
objects younger than a few Myrs is totally meaningless when the age is based on models
using oversimplified initial conditions.} \\
Chabrier et al. (2005) assertain that {\em both models and observations are hampered
by nummerous uncertainties and great caution must be taken when considering young
age ($\le 10$ Myr) objects.}

Chauvin et al. (2004) state in their section 3.5 {\em ... although the models
are reliable for objects with age $\ge 100$ Myr, they are more uncertain
at early phases of evolution ($\le 100$ Myr). As described by Baraffe et al. (2002),
the choice of the initial conditions for the model adds an important source
of uncertainty which is probably larger than the uncertainties associated
with the age and distance of 2M1207.
... We then consider the new generation of models
developed by Chabrier et al. (2000) and Baraffe et al. (2002) ...}
(to determine the mass of 2M1207 b).

\begin{table*}
\begin{tabular}{l|l|l|l}
\multicolumn{4}{c}{\bf Table 4. Summary of parameters for the three planet candidates} \\ \hline
Parameter           & \multicolumn{3}{c}{Objects}\\
                    & GQ Lup b      & 2M1207 b     & AB Pic b    \\ \hline
distance [pc]       & $140 \pm 50$  & $70 \pm 20$: & $47.3 \pm 1.8$ \\ \hline
membership          & Lupus I       & TWA (?)      & TucHorA            \\ \hline
age [Myr]           & $\le 2$       & 5-12:        & 30-40          \\ \hline
epoch difference [yr] & 5           & 1            & 1.5           \\
separation          & 0.7", 100 AU  & 0.8", 54 AU  & 0.5", 258 AU  \\
sign. for CPM (1)  [$\sigma$] & $6+4+7$       & $2+2+4+4$     & $3+5$ \\
remaining motion A/b [mas/yr] & $1.4 \pm 2.2$ & $4.1 \pm 8.2$ & $6.9 \pm 13.2$  \\
orbital motion exp.  [mas/yr] & $3.7 \pm 1.5$ & $1.9 \pm 0.6$ & $6.9 \pm 0.4$  \\
escape velocity exp. [mas/yr] & $5.2 \pm 2.1$ & $2.7 \pm 0.9$ & $9.8 \pm 0.6$ \\
long-term stable ? (2) & yes    &  no           & yes       \\ \hline
SpecType            & M9-L4     &  L5-L9.5      & L0-3      \\
spectrum resolution & 700       &  $< 700$      & 700       \\
spectrum S/N ratio  & 45        &  low          & high      \\
T$_{\rm eff}$ [K]   & 2520-1600 &  2000-1100    & 2400-1600 \\ \hline
gravity $\log g$ [cgs] & 2.0-3.3 (3) & unknown  & unknown   \\ \hline
radius [R$_{\rm jup}$] & $1.2 \pm 0.6$ (4) & unknown & unknown \\ \hline
M$_{\rm K}$ [mag]   & $7.37 \pm 0.96$ & $12.70 \pm 0.75$ & $10.77 \pm 0.14$ \\
$\log L_{bol}/$L$_{\odot}$ & $-2.37 \pm 0.41$ & $-4.49 \pm 0.34$ & $-3.730 \pm 0.039$ \\ \hline
mass [M$_{\rm jup}$] Lyon/Tucson & 1-42 & 2-70        & 11-70    \\
mass [M$_{\rm jup}$] Wuchterl    & 1-3  & n/a (5) & n/a (5)   \\ \hline
\end{tabular}

Remarks: (1) significance for common proper motion in Gaussian $\sigma$;
(2) according to criteria in Weinberg (1987) and Close et al. (2003)
(3) from fit to theoretical GAIA-dusty template spectrum;
(4) from fit to spectrum with flux and temperature known; 
(5) not applicable, because outside of plotted or calculated range. \\
Ref.: this paper, N05a, Mugrauer \& Neuh\"auser 2005, Ch05a, Ch05b, Hipparcos.
\end{table*}

Ch05b write in their section 5 {\em ... as described in Baraffe et al. (2002),
model predictions must be considered carefully as they are still uncertain at early
phases of evolution ($\le 100$ Myrs; see also Mohanty et al. 2004 and Close et al. 2005).
We then considered the most commonly used models of Burrows et al. (1997),
Chabrier et al. (2000), and Baraffe et al. (2002) ...} (to determine the mass of AB Pic b).

It is surprising that Chauvin et al. (2004) and Ch05b first
ascertain that the Lyon (Chabrier et al. 2000
and Baraffe et al. 2002) and Tucson (Burrows et al. 1997) models, which
both do not take into account the collapse and formation, are
not applicable for 2M1207 and AB Pic, and then use them.
Given the fact that these models are not applicable, as correctly
stated by Chauvin et al. (2004) and Ch05b, one has to conclude that
the temperatures and, hence, masses of 2M1207 b and AB Pic b were
essentially undetermined.

The model by Wuchterl \& Tscharnuter (2003) for stars and brown dwarfs
does take into account their formation, so that it can be valid for 
very young objects. 
The tracks for planets shown in Fig. 4 in N05a are calculated based
on the nucleated instability hypothesis (Wuchterl et al. 2000).

Finally, we would like to point out, that neither distance nor age are
directly derived parameters in the cases of the companions,
and that only the distance towards AB Pic A is determined directly
as parallaxe by Hipparcos. In all three cases, the age and distance of
the companion is assumed to be the same as for the primary because
of common proper motion. However, there are counter-examples.

\section{Summary and discussion on planethood}

We compile all information relevant for our discussion in table 4.

{\bf Gravitationally bound~?}
While the remaining possible motion between GQ Lup A and b (change in
separation and position angle) is smaller than both the expected 
orbital motion and the expected escape velocity, this system may well
be gravitationally bound. 
This may be different for 2M1207 A+b: The remaining
motion between the two objects may be larger
than the expected escape velocity (table 4), so that it
is not yet shown to be bound. The GQ Lup system has a total mass and
bounding energy sufficient for being long-term stable according to
the criteria by Weinberg et al. (1987) and Close et al. (2003), 
while the 2M1207 system is not -- too low in mass(es) for the given separation.
See Mugrauer \& Neuh\"auser (2005) for a discussion.
2M1207 A+b, if formed together and if still young, may be
an interesting case as a low-mass binary just desintegrating.
The remaining motion between the AB Pic A and b is not yet shown to 
be smaller than the expected escape velocity (table 4),
so that is not yet shown to be bound, but it could be stable.

{\bf Masses:}
Chauvin et al. (2004) and Ch05b may have underestimated the 
range in possible temperatures of both
2M1207 b and AB Pic b by using models
rather than spectral type to temperature conversions.

According to the Lyon and Tucson models, GQ Lup b and 2M1207 b may either be
planet or brown dwarf, while AB Pic b would be a low-mass brown dwarf.
According to both the Wuchterl model and our K-band spectrum compared
with the Hauschildt GAIA-dusty model, GQ Lup b is a planet;
the mass, age, and planethood of AB Pic b and 2M1207 b cannot yet be
discussed using the Wuchterl model, because it is not yet available
in the neccessary parameter range regarding temperatures, 
luminosities, and ages (outside the range in fig. 4 in N05a).
An extrapolation would indicate that AB Pic b has a mass
around one Jupiter mass, but it is probably much harder to form a
one Jupiter mass object at 258 AU separation (AB Pic b)
than at 100 AU separation (GQ Lup b).
2M1207 A, according to the Wuchterl model, appears to be below the 
D burning mass limit at roughly a few Jupiter masses. 
It would need to be more nearby and/or older to be above 13~M$_{\rm jup}$.

Lodato et al. (2005) argue that 2M1207 system may rather be seen
as a binary pair of two very low-mass objects than a planet around a primary,
due to the similar masses of both objects at the given relatively large separation.
A several Jupiter mass object at 55 AU separation could not have formed
by planet-like core accretion from a (low-mass) disk around a brown 
dwarf such as 2M1207 A. 
GQ Lup b may have been able to form as a planet at 100 AU 
separation around an almost solar-mass star (Lodato et al. 2005),
while AB Pic b with 258 AU separation is again to far off.

The GQ Lup b K-band spectrum resembles well the spectra of isolated, 
free-floating young objects previously classified as brown dwarfs.
However, the mass range of brown dwarfs and planets may well overlap:
The Saturnian moon Titan has both a solid core and an atmosphere.
It is regarded as a moon, because it orbits Saturn, a planet.
If Titan would orbit the Sun directly (without other objects
of similar mass in a similar orbit), it would be regarded a planet.
The mass range of moons and planets overlaps.
Analogously, if an object below 13 Jupiter masses with a solid or fluid
core formed in a disk and orbits a star, 
then its a planet, otherwise a low-mass brown dwarf.

Those young objects with similar spectra as GQ Lup b were classified as 
13 to 78 Jupiter mass brown dwarfs based on Tucson or Lyon models,
which may not be valid until at least 10 Myrs, as recently specified in
Chabrier et al. (2005). Hence, they may be lower in mass,
maybe around a few to 10 Jupiter masses.
If GQ Lup b has the same spectrum and mass as those young free-floating
objects, the free-floating objects may be low-mass brown dwarfs
(or planemos),
because they are free-floating, while GQ Lup b with the same mass can 
be a planet, namely if formed with a core in a disk.

{}

%
%
%
%
%
%
%
%
%
%


\end{document}